\documentclass[aps,prl,reprint,amssymb,longbibliography,preprintnumbers,twocolumn,amsmath,superscriptaddress,floatfix]{revtex4-2}
\usepackage{graphicx}
\usepackage{bm}
\usepackage{hyperref}
\usepackage{physics}
\usepackage{orcidlink}
\usepackage[final]{showlabels}

\usepackage{color}
\usepackage[normalem]{ulem}

\newcommand{\be}{\begin{equation}}
	\newcommand{\ee}{\end{equation}}
\newcommand{\ba}{\begin{eqnarray}}
	\newcommand{\ea}{\end{eqnarray}}
\newcommand{\bs}{\begin{subequations}}
	\newcommand{\es}{\end{subequations}}
\newcommand{\bw}{\begin{widetext}}
	\newcommand{\ew}{\end{widetext}}

\begin{document}

\title{Beyond Electric-Dipole Treatment of Light-Matter Interactions in Materials: Nondipole Harmonic Generation in Bulk Si}

\author{Simon Vendelbo Bylling Jensen \orcidlink{0000-0002-6749-0978}}
\affiliation{Department of Physics and Astronomy, Aarhus
University, DK-8000 Aarhus C, Denmark}
\affiliation{Max Planck Institute for the Structure and Dynamics of Matter and Center for Free-Electron Laser Science, Hamburg 22761, Germany}

\author{Nicolas Tancogne-Dejean \orcidlink{0000-0003-1383-4824}}
\affiliation{Max Planck Institute for the Structure and Dynamics of Matter and Center for Free-Electron Laser Science, Hamburg 22761, Germany}

\author{Angel Rubio \orcidlink{0000-0003-2060-3151}}
\affiliation{Max Planck Institute for the Structure and Dynamics of Matter and Center for Free-Electron Laser Science, Hamburg 22761, Germany}
\affiliation{Center for Computational Quantum Physics (CCQ), The Flatiron Institute, New York, New York 10010, USA}

\author{Lars Bojer Madsen \orcidlink{0000-0001-7403-2070}}
\affiliation{Department of Physics and Astronomy, Aarhus
University, DK-8000 Aarhus C, Denmark}

\date{\today}

\begin{abstract}
A beyond electric-dipole light-matter theory is needed to describe emerging X-ray and THz applications for characterization and control of quantum materials but inaccessible as nondipole lattice-aperiodic terms impede on the use of Bloch's theorem. To circumvent this, we derive a formalism that captures dominant nondipole effects in intense electromagnetic fields while conserving lattice translational symmetry. Our approach enables the first accurate nondipole first-principles microscopic simulation of nonperturbative harmonic generation in Si. We reveal nondipole-induced transverse currents generating perturbative even-ordered harmonics and display the onset of nondipole high harmonic generation near the laser damage threshold.

\end{abstract}  
\maketitle

\begin{figure*} 
	\includegraphics[width=17.2 cm]{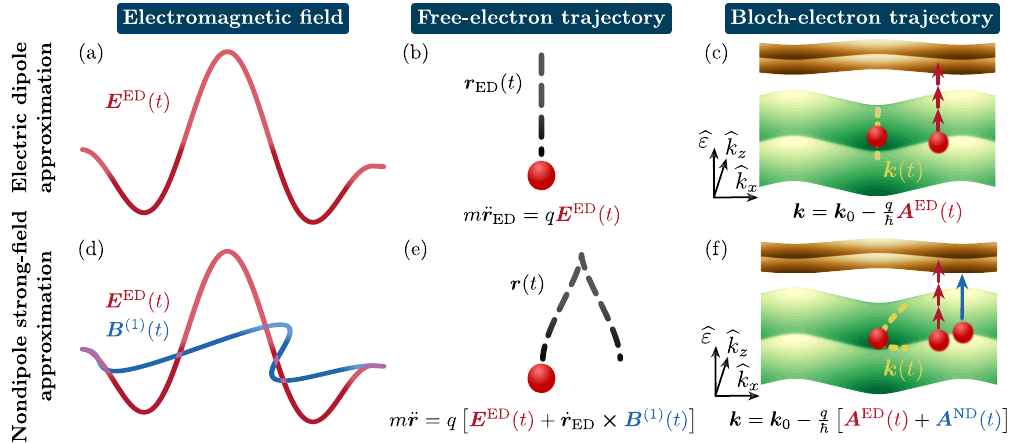}
	\caption{Illustration of the ED (a-c) and nondipole strong-field (d-e) methodologies. For (a) and (d), the electromagnetic field is illustrated with the ED electric field $\bm E^{\text{ED}}(t)$ and leading-order nonvanishing magnetic field component $\bm B^{(1)}(t)$. The trajectories induced by the fields of (a) and (d) are illustrated for a free electron in (b) and (d) alongside the associated equations of motion. Similarly, the field-induced intra- and interband Bloch-electron dynamics are illustrated as momentum-space trajectories in (c) and (f) alongside their explicit crystal momentum expressions following from the acceleration theorem.} \label{fig:1}
\end{figure*}

The electric dipole approximation (EDA) forms a cornerstone in our fundamental understanding of light-matter interactions. Textbook examples are selection rules for transitions between quantum states in atomic and molecular physics~\cite{bransden2003physics}, giving rise to spectroscopic principles~\cite{TF9262100536,PhysRev.28.1182} and the dominance of direct interband transitions that form the fundamental processes of photonics and material science~\cite{Ashcroft76}. In its simplicity, the EDA consists of approximating the propagation factor of an electromagnetic field as $\exp(i\bm k_L \cdot\bm r) \simeq 1$, with wave vector of size $k_L = \omega/c$ in the light propagation direction, position $\bm r$, frequency $\omega$ and speed of light~$c$. By neglecting spatial variations of the field, the EDA allows for analytical understanding and efficient first-principles simulations. Its validity is often assured as $i\bm k_L \cdot\bm r \ll 1 $ due to the relative confinement of an atomic or molecular system compared to the electromagnetic wavelength, or due to its negligible wave vector compared to the range of crystal momenta within a Brillouin zone. 

Intuitively, the EDA breaks down for large $k_L$, which occurs as the wavelength subceed the atomic scale of $\lambda = 1 $ a.u.~\cite{Ilchen2018}.
Seemingly counter-intuitively, a breakdown also occurs in the opposite long-wavelength regime, if the electromagnetic field is of high intensity, $I$ exceeding $I/(8mc\omega^3) \sim 1$ a.u. Here the laser-induced quiver motion of a charged particle, scaling with $E_0/\omega^2$, grants a non-negligible beyond-dipole $\dot{\bm r} \cross \bm B$ magnetic field interaction from the Lorentz force~\cite{PhysRevLett.101.043002,Reiss_2014,PhysRevLett.113.243001} and eventually leads to relativistic effects~\cite{PhysRevA.110.033108}.
Despite breakdowns in both wavelength limits, the EDA is accurate for the regimes of historical interest for spectroscopy being soft X-ray, ultraviolet, visible, and near-IR.

However, emerging light sources and interest in fundamental science and technological applications beyond these wavelength and intensity regimes highlight the necessity to extend our theoretical capabilities to exceed beyond an electric dipole (ED) description. An example of such intrinsically beyond-ED applications is the use of tailored macroscopic field structures such as orbital angular momentum vortex beams for advancements in quantum information science, microscopy, and spectroscopy~\cite{Habibović2024,Gauthier2017,Forbes2024}. The need for beyond-ED descriptions is especially evident for condensed matter applications, where, however, they are hindered by fundamental restrictions from Bloch's theorem. Bloch's theorem relies on the periodic translational symmetry of the underlying potential $V(\bm r)=V(\bm r + \bm R)$ with translation vector $\bm R$ defined from the crystal lattice constant. Such periodic symmetry considerations are incompatible with nondipole multipole expansions of the electromagnetic field, as these introduce nonperiodic $\bm r$-dependencies that dynamically couple electronic states of different crystal momenta. The nondipole-induced breakdown of Bloch's theorem thus couples all crystal momenta and makes it almost impossible for first-principles light-matter modeling to exceed beyond the dipolar optical regime, so far confining it to density functional perturbation theory~\cite{PhysRevLett.91.196401}.

In the single-photon ionization regime, nondipole light-matter interactions have been observed from condensed matter systems for almost a decade and arise as a distinct forward-backward anisotropy within angle-resolved photoelectron spectroscopy (ARPES)~\cite{refId0,HEMMERS2004123}. In the opposite long-wavelength THz regime, nondipole effects remain unexplored but are expected to participate in the developing fields of lightwave electronics~\cite{Borsch2023,Reimann2018}, orbital angular momentum light-driven applications \cite{Sederberg2020}
and quantum material science~\cite{RevModPhys.93.041002}. With imminent interest in such research fields, and ARPES being the leading tool for investigation of light-dressed quantum systems~\cite{PhysRevResearch.4.033101,Reimann2018}, a nondipole condensed matter treatment is crucial to advance our understanding from static perturbative models to accurate first-principles dynamical simulations.
Another timely topic where nondipole effects are expected to participate is the self-probing mechanism of high-order harmonic generation (HHG), which often operates in the IR strong-field regime.
Historically, HHG is utilized to generate intense ultrashort laser pulses for accommodating time-resolved measurements of electron processes~\cite{Rupp2017,doi:10.1126/science.1218497,doi:10.1126/science.1157846,Tzallas2011,PhysRevLett.103.028104,PhysRevLett.105.093901},
but HHG also allows for revealing characteristics of the underlying ultrafast electron processes within the generating media~\cite{PhysRevA.66.023805,PhysRevLett.98.203007,doi:10.1126/science.1163077,doi:10.1126/science.1123904,PhysRevLett.94.053004,Itatani2004}. 
In this setting, HHG spectroscopy is utilized within condensed matter systems for reconstruction of electronic bonds~\cite{Lakhotia2020}, bandstructures~\cite{PhysRevLett.115.193603}, Berry curvatures~\cite{Luu2018}, phonon dynamics~\cite{Zhang2024}, and phase transitions~\cite{PhysRevResearch.3.023250}. The predicted sensitivity towards topology~\cite{PhysRevLett.130.166903,PhysRevX.13.031011}, intraband processes~\cite{Ghimire2011}, quasi-particles~\cite{PhysRevA.109.063104,doi:10.1126/sciadv.adn6985}, strong correlations~\cite{PhysRevLett.121.097402,Silva2018} and twist angles~\cite{Molinero2024} offers great opportunities for ultrafast all-optical spectroscopic applications. 
Sensitivity towards nondipole magnetic field interactions is also expected, and known to diminish the efficiency of HHG in gaseous media, as nondipole radiation pressure perturbs the ionized electron trajectory, such that it misses its parent ion for the recombination step~\cite{PhysRevLett.85.5082}. In solids, however, nondipole descriptions remain of semiclassical nature~\cite{PhysRevA.105.L021101} due to first principles simulations being limited by the symmetry restrictions of Bloch's theorem.

Here, we provide a first-principles formalism, which is simple to implement and accounts for leading-order nondipole light-matter interactions in the strong-field regime, without violating periodic symmetry considerations. We apply it to elucidate the role of nondipolar interactions for harmonic generation in crystalline Si. We obtain even-ordered perturbative nondipole harmonics, the first sign of nondipole effects from solids in the long-wavelength regime. We further predict the onset of nonperturbative high-order nondipole harmonics for field intensities approaching the material damage threshold. In experiments, the appearance of such symmetry-forbidden harmonics is known to occur if approaching the material damage threshold but is commonly justified as the first sign of lattice aperiodicity induced by laser damage, or charge polarization from the photo-Dember effect \cite{dember1931photoelektronische,PhysRevB.109.L140304}. We allude to nondipolar light-matter interactions playing a significant role in such observations. 

To obtain the condensed-matter nondipole strong-field methodology, we consider a general spatially dependent external electromagnetic field with vector potential $\bm{A}(\bm r, t)$, which is not restricted to be of commensurate periodicity with any underlying lattice but can take any form. In general, the field can be composed of multiple laser pulses, $\bm A (\bm r, t) = \sum_i 	\bm A_i (\omega t + \chi_i)$, with polarization planes perpendicular to their propagation directions $\widehat{\bm k}_{L,i}$, such that $\chi_i= -\omega \bm r \cdot \widehat{\bm k}_{L,i} /c$. A Taylor expansion of $\bm A (\bm r, t) = \sum_{l=0}^\infty \bm A^{(l)}$ around $\chi_i = 0$ leaves a multipolar series of vector potentials with superscript $(l)$ to highlight their orders of $\chi_i^{(l)}$. The electric field, $\bm E^{(l)} = - \partial_t \bm A^{(l)}$, and magnetic field, $\bm B^{(l)}=\bm \nabla \times \bm A^{(l)}$, inherits the superscript of the associated vector potential. %Note that the fields are considered in Coulomb gauge and neglecting induced electrostatic potentials. 
The zeroth-order term is simply the ED field $\bm A^{\text{ED}} = \bm A^{(0)} (t) = \sum_i \bm A_i (\omega t + \chi_i)|_{\chi_i= 0}$. As illustrated in Fig.~\ref{fig:1}~(a) the ED field consists of only an electric field, $\bm E^{\text{ED}} = \bm E^{(0)} = \sum_i \bm E^{\text{ED}}_i$, with $\bm E^{\text{ED}}_i = - \partial_t \bm A^{(0)}_i$, and no magnetic field components as $\bm B^{\text{ED}}= \bm B^{(0)}=\bm \nabla \times \bm A^{(0)} = 0$. For a charged particle, as illustrated in Fig.~\ref{fig:1}~(b), the electric field of Fig.~\ref{fig:1}~(a) induces a trajectory, $\bm r_{\text{ED}}(t)$, as prescribed by the associated ED Lorentz force. Within a solid, such spatially independent field can fulfill the symmetries of Bloch's theorem in a velocity-gauge description and can therefore induce a Bloch electron trajectory $\bm k_D (t)$, as illustrated in Fig.~\ref{fig:1}~(c), prescribed by the acceleration theorem. To go beyond the EDA might, at first sight, be contradictory with the symmetry requirements of Bloch's theorem, as the first nondipole correction, $\bm A^{(1)}$, contains linear spatial dependencies. However, periodicity can be achieved if utilizing the nondipole strong-field-approximation Hamiltonian formalism developed for atoms and molecules~\cite{PhysRevA.101.043408}. For a free electron, this formalism accounts for the effect of the leading-order magnetic field correction $\bm B^{(1)}$ of Fig.~\ref{fig:1}~(d), when acting on the ED trajectory $\bm r_{\text{ED}}$ of Fig.~\ref{fig:1}~(b). The approach breaks forward-backward symmetry by introducing a magnetic-field-induced radiation pressure as illustrated in Fig.~\ref{fig:1}~(e). Extending the nondipole strong-field-approximation methodology to condensed-matter systems brings about a spatially independent nondipole correction $\bm A^{\text{ND}}$ to the electromagnetic vector potential. It fulfills Bloch's theorem and alters both the intraband and interband Bloch electron trajectories as depicted in Fig.~\ref{fig:1}~(f). 

To arrive at the associated Hamiltonian, we consider minimal coupling $\bm p \rightarrow \bm p - q \bm A (\bm r, t) $ for a charged particle in an effective potential $V(\bm r)=V(\bm r + \bm R)$, which can arise from heavier nuclei along with mean-field electron interactions as typically described in density functional theory approaches. Expanding to first order of $\chi_i$ corresponds to including terms of $\bm A^{\text{ED}}$ and $\bm A^{(1)}$ and grants the Hamiltonian
\be
H=  \frac{[\bm{p} - q  \bm A^{\text{ED}}]^2}{2m} - \frac{q}{m} \bm A^{(1)} \cdot \bm p  + \frac{q^2}{m} \bm A^{\text{ED}} \cdot \bm A^{(1)}+ V.
\label{Happ11}
\ee
The nondipole strong-field approximation now consists of neglecting the term $(-q/m) \bm A^{(1)} \cdot\bm {\bm p}$ compared with the $ (q^2/m )\bm A^{\text{ED}} \cdot \bm A^{(1)}$ term and corresponds to accounting only for the leading-order magnetic field correction on the ED trajectory as illustrated in Fig.~\ref{fig:1}~(e) and discussed in detail in Ref.~\cite{PhysRevA.101.043408}. It is valid in the strong-field regime where the approach surpasses the accuracy of a conventional $\bm A^{(1)}$ multipolar expansion by circumventing numerical instabilities. 
Rewriting $\bm A^{(1)} = \sum_i \bm E^{\rm ED}_i \left(\bm r \cdot \widehat{\bm k}_{L,i}\right)/ c$, we consider a unitary transformation with $U=\exp \left(\frac{i}{\hbar} q \bm A^{\text{ND}} \cdot \bm r \right)$ leading to the nondipole (ND) strong-field-approximation (SFA) Hamiltonian
\be
H_\text{ND}^{\text{SFA}}= \frac{\left[\bm{p} - q  \left(\bm A^{\text{ED}}  + \bm A^{\text{ND}}\right) \right]^2}{2m} + V,
\label{NDSFAVG1}
\ee
with 
\be
\bm A^{\text{ND}} = \frac{q}{mc} \int dt  \sum_i \left(\bm A^{\text{ED}} \cdot \bm E^{\text{ED}}_i\right)  \widehat{\bm k}_{L,i}.
\label{AMlongpulse1}
\ee
The obtained nondipole strong-field-approximation Hamiltonian allows for extending any ED theory by simply correcting the electromagnetic vector potential $ \bm A^{\text{ED}}$ with its associated nondipole correction $\bm A^{\text{ND}}$ defined only from the components of $ \bm A^{\text{ED}}$ itself along with their propagation directions $\widehat{\bm k}_{L,i}$. The formalism can thus be readily implemented in any condensed matter or quantum chemistry theoretical framework at negligible computational cost and without requiring nonlocal pseudopotentials~\cite{PhysRevLett.87.087402,PhysRevLett.91.196401}. For a free electron, the approach provides the equation of motion as sketched in Fig.~\ref{fig:1}~(e)~\cite{PhysRevA.101.043408}. The Hamiltonian of Eq.~\eqref{NDSFAVG1} fulfills Bloch's theorem by only containing momentum operators and not inflicting with the spatial periodicity of the effective potential. It thus captures nondipole interactions in strong-field driven condensed-matter systems and modifies the inter- and intraband electron dynamics as sketched in Fig.~\ref{fig:1}(f), by introducing a nondipole correction of Eq.~\eqref{AMlongpulse1} oscillating with twice the laser frequency.
The intraband nondipole correction introduces a transverse current along $\widehat{\bm k}_{L,i}$, reminiscent of the transverse currents induced by Berry curvature or spin-orbit effects~\cite{PhysRevX.13.031011,PhysRevLett.130.166903}. The significance of the nondipole intraband trajectory deflection scales with $A^{\text{ND}}/A^{\text{ED}} \sim q\sqrt{I}/(2m \omega c)$, causing a percentage-wise deviation for Ti:sapphire  laser ($\lambda = 800$~nm) intensities approaching $1000$~TW/cm$^2$ or for CO$_2$ laser ($\lambda = 10.6 $~$\mu$m) intensities of $5$~TW/cm$^2$. The physical consequences of the intraband nondipole transverse current strongly depend on the material properties and can have a crucial role for electrons driven near critical points of the bandstructure. Nondipole modifications to the interband electron dynamics are caused by $2\omega$-transitions whose impact will strongly depend on the details of the material bandstructure.

The proposed nondipole treatment is expected to be accurate based on substantial precedents from atomic strong-field physics~\cite{PhysRevA.48.R4027,PhysRevA.64.013411,PhysRevA.93.013423,PhysRevA.101.043408}, where it has proven fruitful for analytical descriptions of laser-assisted scattering~\cite{Jensen2020}, laser-induced adiabatic states~\cite{Madsen2021}, above-threshold ionization~\cite{Lund_2021,PhysRevA.106.043118,PhysRevA.110.023111}, and tunneling ionization~\cite{PhysRevA.105.043107}. Note that a further unitary transformation of the Hamiltonian in Eq.~\eqref{NDSFAVG1}, as described in Ref.~\cite{PhysRevA.101.043408}, may express the interaction in terms of $ \widetilde{\bm E} \cdot \bm r$ for a suitably modified field $\widetilde{\bm E}$. We shall not pursue this formalism here as the presence of $\bm r$ breaks the $\bm R$-periodicity of $V(\bm r)$.

\begin{figure} 
	\includegraphics[width=\columnwidth]{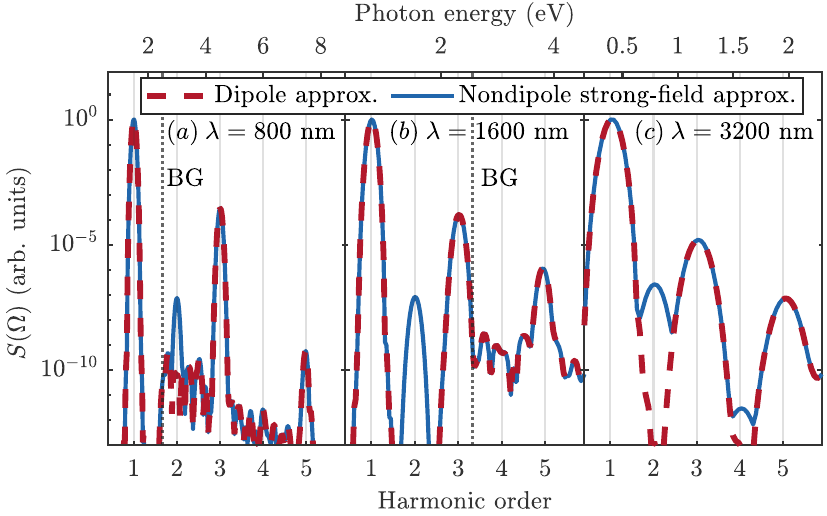}
	\caption{(a)-(c) HHG spectra from Si for driving laser wavelengths of $800$, $1600$, and $3200$ nm, respectively. The spectra are obtained both within the EDA and beyond with laser and system parameters given in the text. The dotted line labeled by BG shows the bandgap energy in units of harmonic order and remains out of scope in (c).} \label{fig:2}
\end{figure}

\begin{figure} 
	\includegraphics[width=\columnwidth]{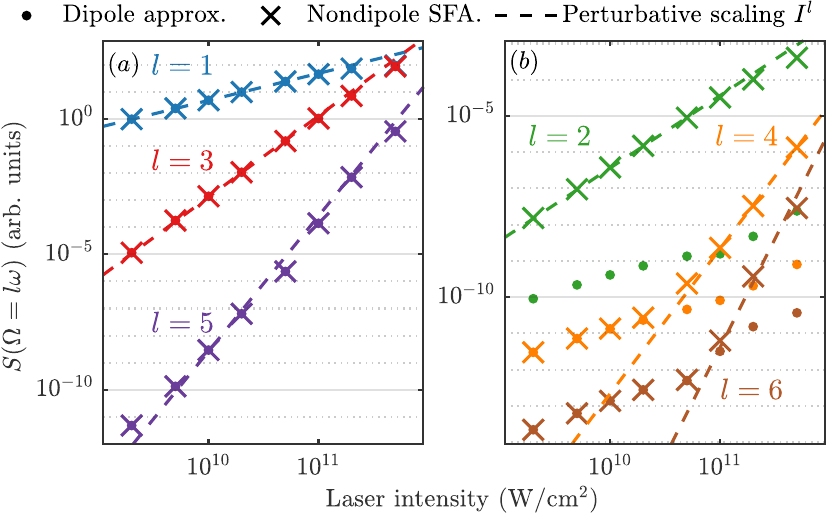}
	\caption{Emitted harmonic intensity as a function of driving laser intensity for odd and even harmonic orders of Si, respectively, in (a) and (b). The EDA and nondipole strong-field-approximation Hamiltonian (Nondipole SFA) light-matter formalism is considered for a $800$ nm laser field with field and system parameters as of Fig. \ref{fig:2}~(a). For the even harmonic orders of (b) the EDA predictions are indistinguishable from the numerical noise floor.} \label{fig:3}
\end{figure}

To identify spectroscopic characteristics of beyond-ED interactions in condensed-matter HHG, we consider the nondipole strong-field-approximation Hamiltonian within time-dependent density-functional theory (see End Matter). Harmonic spectra from Si driven with three different laser wavelengths are given in Fig.~\ref{fig:2} and display the occurrence of nondipole even-ordered harmonics, which would otherwise be forbidden by the inversion symmetry of the sample material. For Fig.~\ref{fig:2}~(a) the second-order above-bandgap harmonic suggests the appearance of an additional nondipole interband transition pathway, as illustrated in Fig.~\ref{fig:1}~(f). The interband transition probability, enhanced by the increasing density of states near the bandgap, provides a significant nondipole second-order harmonic when compared with the fifth-order harmonic. At longer wavelengths in Fig.~\ref{fig:2}~(b), the nondipole harmonic is now of below-bandgap intraband nature and attributed to the transverse currents induced by the modified intraband electron trajectory as illustrated in Fig.~\ref{fig:1}~(f). Comparing Fig.~\ref{fig:2}~(b) to (c) we observe the intraband nondipole harmonics to have a relative increase of a factor of $3.20$ when doubling the wavelength. A scaling, which fits relatively well with what is obtained from solving the semiclassical intraband model within the nondipole strong-field-approximation Hamiltonian formalism (see End Matter), which in the perturbative regime approach a scaling of $S(2\omega)/S(\omega) \propto I \lambda^2$. 
In Fig.~\ref{fig:2}~(c) one can similarly see the onset of a fourth-order nondipole harmonic. To investigate the appearance of such higher-order nondipole harmonics, the emission of different harmonic orders is examined for varying driving laser intensities at $800$ nm in Fig.~\ref{fig:3}.


All odd harmonic orders, displayed in Fig.~\ref{fig:3}~(a), are captured well within both the ED and the nondipole approach. The lowest harmonic orders attain near perturbative scaling of $S \propto I^l$ for the $l$'th harmonic order, but the higher-order processes deviate slightly from this trend due to the nonperturbative nature of the underlying strong-field process and band-structure nonlinearities. The even-ordered harmonics are displayed in Fig.~\ref{fig:3}~(b) and are only captured within the nondipole light-matter description as the ED signal is indistinguishable from numerical noise. A perturbative second harmonic is observed as well as the appearance of a perturbative fourth harmonic and nonperturbative sixth harmonic in the high-intensity regime. The highest intensities displayed in Fig.~\ref{fig:3} give rise to a valence electron excitation %of $5 \%$, 
approaching the $10 \%$ excitation threshold for material damage of Si within the electron-hole plasma model \cite{PhysRevB.42.7163}.
From our data, we can extract one component of the effective bulk second-order susceptibility $\chi^{(2)}$ (see End Matter). In the low-intensity limit, we obtain $\chi^{(2)} = 0.54 \times 10^{-18}$~m$^2$/V, which is comparable with the experimental anisotropic bulk quadrupolar susceptibility for silicon~\cite{footnote}, $\zeta = 0.23 \times 10^{-18}$~m$^2$/V and $4.4 \times 10^{-18}$~m$^2$/V of Refs. \cite{FALASCONI2001105,PhysRevB.75.241307}.
For the given laser parameters, a time-frequency analysis reveals that the additional interband nondipole resonances illustrated in Fig.~\ref{fig:1}(f), have negligible impact on the excitation degree and the recombination dynamics \cite{PhysRevLett.124.153204}. However, such additional excitation pathways could in other laser regimes impact photo-conductivity and play a competing role in resonant second harmonic spectroscopies \cite{PhysRevLett.129.227401}. Based on our findings, nondipole harmonics are observed to be a robust feature of strong-field-driven materials and are expected to influence the laser-driven nonperturbative electron dynamics when approaching the material damage threshold. The nondipole transverse deviation of the intraband electron trajectories could significantly alter the electron effective mass and associated dynamics for specific material orientations and could thus also affect the intensity of odd ordered harmonics. HHG spectroscopy is thus expected to be sensitive to nondipole effects, which can provide features of symmetry breaking and induce a transverse current similar to topology or spin-orbit coupling~\cite{PhysRevX.13.031011}. Such features might, however, be distinguished by characteristic polarization properties as the generated transverse nondipole current tends to polarize along the laser propagation direction. Semiclassical modeling corroborates such polarization relations and selection rules and predicts a nondipole HHG spectral plateau in the long-wavelength strong-field regime~\cite{PhysRevA.105.L021101}. Such predictions can be readily elucidated by combining the first-principles methodology at hand with experimental efforts, where distinctive measurements of beyond-dipole harmonics are expected to require thin-sample geometries or multi-excitation measurements to assure phase-matching, alike schemes for measuring quadrupolar quantum-optical HHG features~\cite{Gorlach2020}.

In summary, we derived a methodology for accurately accounting for leading-order magnetic field effects in strong-field-driven condensed-matter systems. We perform what, to our knowledge, constitutes the first simulations from first principles of nondipole electron dynamics in a condensed matter system for elucidating nondipole features of harmonic generation from Si. We find a nondipole-induced transverse current that generates even-ordered harmonics whose magnitude reliably reproduces experimental susceptibilities and is expected to play a role in ultrafast spectroscopies exceeding the optical regime. The formalism predicts both perturbative scaling low-order nondipole harmonics, but also higher-order nonperturbative nondipole harmonics, which are expected from semiclassical models to form a spectral plateau in the strong field limit~\cite{PhysRevA.105.L021101}. Whether such a nondipole HHG mechanism follows identical cutoff scaling as the ED case~\cite{PhysRevA.109.063109}, is a topic for future investigations. The separate nature of inter- and intraband nondipole dynamics can also be scrutinized if implementing the nondipole formalism within the widely used semiconductor Bloch equation-formalism~\cite{madsmaster}. While we focused here on cubic Si, a more complex interplay due to nondipole fields is expected for noncubic materials. 

The derived first-principles formalism provides a simple protocol to extend any ED-approximated theory to include beyond-ED couplings at negligible computational cost. To this end, we highlight that the formalism will be available as a feature in future releases of the Octopus multisystem simulation framework~\cite{10.1063/1.5142502}. It inspires an array of interesting investigations to elucidate nondipole influences in research areas such as excitons~\cite{Koch2006,PhysRevA.109.063104}, carrier mobility~\cite{10.1063/1.1394953,10.1063/1.2980026,Su:09}, photoconductivity~\cite{PhysRevB.62.15764}, impact ionization~\cite{PhysRevB.79.161201}, phonon-plasmon systems~\cite{PhysRevLett.94.027401} or polaron dynamics~\cite{Gaal2007}. Although the formalism is formulated based on time-dependent electromagnetic fields, it can also capture certain electrostatic field effects and hence opens up possibilities for microscopic \textit{ab initio} modeling of processes in static electric and magnetic fields.

\begin{acknowledgments}
This work was supported by the Danish Council for Independent Research (GrantNo.9040-00001B). S.V.B.J. further acknowledges support from the Danish Ministry of Higher Education and Science and the Max Planck Society. Fruitful discussions with F. P. Bonafé are acknowledged. 
\end{acknowledgments}

%\bibliography{references}

\begin{thebibliography}{84}
\bibitem [{foo()}]{footnote}%
	\BibitemOpen
	\href@noop {} {\bibinfo  {journal} {In cubic materials, $\zeta$ is given in
			terms of the components of the susceptibility by $\zeta =
			\chi^{(2)}_{\mathrm{bulk},iiii} - \chi^{(2)}_{\mathrm{bulk},ijij} -
			\chi^{(2)}_{\mathrm{bulk},ijji} - \chi^{(2)}_{\mathrm{bulk},iijj}$, see Ref.
			[63]}\ }\BibitemShut {NoStop}%
\end{thebibliography}
%\input{ms.bbl}

\onecolumngrid

\subsection{End Matter}
\twocolumngrid
\textit{Numerical methods} 
\noindent
First-principles simulations are performed of a Si crystal utilizing real-space, real-time, time-dependent density-functional theory with the Octopus software~\cite{10.1063/1.5142502}. The ground state orbitals are obtained by iterative diagonalization from a $28$ x $28$ x $28$, four times shifted Monkhorst-Pack grid~\cite{PhysRevB.13.5188}. Simulations employ a $10.26$ a.u. lattice constant, a $0.50$ a.u. spatial and $0.1$ temporal grid spacing and the adiabatic local density approximation~\cite{PhysRev.140.A1133}. Electrons are driven along the $\overline{\Gamma X}$ similar to Ref.~\cite{Tancogne-Dejean2017}. The considered sin$^2$ driving electromagnetic field attains a duration of $100$ fs, a peak electric field strength of $10^{10}$ W/cm$^2$ and are considered with wavelengths of $800$, $1600$ and $3200$ nm. The emitted HHG spectrum is obtained from the Fourier transformed current, $S(\Omega) \propto \abs{\Omega j(\Omega)}^2$, with the application of a cos$^8$ Fourier window function centered around the driving field maxima to remove numerical artifacts.

\textit{Optical susceptibility}
\noindent
The second-order susceptibility is obtained under the assumption of a quasimonochromatic applied field $E_z (\omega)$, which is sharply peaked at the driving frequency $\omega$. This allows to relate the second-order polarizability $P^{(2)}_{xzz}$ to the susceptibility as \cite{10.1063/1.2790014,PhysRevLett.121.097402}
\begin{equation}
\chi^{(2)}(-2\omega,\omega,\omega)_{x,z,z} = \frac{ P^{(2)}_{xzz}(2 \omega)}{\int_{-\omega_c}^{\omega_c} \frac{d\omega'}{2\pi} E_z(\omega - \omega') E_z(\omega + \omega')}.  \nonumber
\end{equation}
Here, assuming negligible intraband currents, the polarizability can be extracted from the current as
\begin{equation}
P^{(2)}_{xzz}(2 \omega) =  -\frac{i}{2\omega} j^{(2)}_{x}(2\omega).  \nonumber
\end{equation}
The second-order susceptibility is retrieved and compared to experimental values in the main text.

\textit{Semiclassical intraband model}
\noindent
Scaling relations are extracted for a localized electron wavepacket at $\bm r = 0$ and $\bm k = 0$ propagated through the semiclassical intraband model~\cite{PhysRevB.59.14915,Luu2015,PhysRevA.105.L021101}. Utilizing the strong-field-approximation Hamiltonian formalism the equations of motion become
\begin{equation}
\hbar \dot{\bm k} = q\left(  - \partial_t  \left(\bm A^{\text{ED}}  + \bm A^{\text{ND}}\right) \right) \ \   \mathrm{and}  \ \ \   \dot{\bm r} = \frac{1}{\hbar}\pdv{\varepsilon(\bm k)}{\bm k}, \label{rdot}
\end{equation}
with bandstructure of $\varepsilon(\bm k)$. An analytical solution in the perturbative limit can be obtained using the asymptotic form of the Bessel function $\mathrm{J}_l(x) \sim \left[\Gamma (l + 1)\right]^{-1}\left(x/2\right)^l$. Doing so provides a significance of the emitted nondipolar harmonics which scales as $S(2\omega)/S(\omega) \propto I \lambda^2$ as also given in the main text.

\end{document}